\documentclass[twocolumn, trackchanges]{aastex701}

\usepackage{graphicx, caption, subcaption, rotating}

\usepackage{amsmath}
\begin{document}

\shortauthors{Johnstone et al.}
\shorttitle{Radio-Near IR Imaging of Dual AGN Candidates}

\title{Radio-Near Infrared Imaging of Dual Active Galactic Nuclei Candidates}

\author[0000-0001-7690-3976]{Makoto A. Johnstone}
\affiliation{Department of Astronomy, University of Virginia, 530 McCormick Road, Charlottesville, VA 22903, USA}
\email{fhh3kp@virginia.edu}

\author[0000-0001-9163-0064]{Ilsang Yoon}
\affil{National Radio Astronomy Observatory, 520 Edgemont Road, Charlottesville, VA 22903, USA}
\affiliation{Department of Astronomy, University of Virginia, 530 McCormick Road, Charlottesville, VA 22903, USA}
\email{iyoon@nrao.edu}

\author[0000-0003-3168-5922]{Emmanuel Momjian}
\affiliation{National Radio Astronomy Observatory, P.O. Box O, Socorro, NM 87801, USA}
\email{}

\author[0000-0003-0057-8892]{Loreto Barcos-Munoz}
\affiliation{National Radio Astronomy Observatory, 520 Edgemont Road, Charlottesville, VA 22903, USA}
\affiliation{Department of Astronomy, University of Virginia, 530 McCormick Road, Charlottesville, VA 22903, USA}
\email{lbarcos@nrao.edu}

\author[0000-0003-2638-1334]{A.S. Evans}
\affiliation{Department of Astronomy, University of Virginia, 530 McCormick Road, Charlottesville, VA 22903, USA}
\affiliation{National Radio Astronomy Observatory, 520 Edgemont Road, Charlottesville, VA 22903, USA}
\email{aevans@nrao.edu}

\author[0000-0003-2983-815X]{Bjorn Emonts}
\affiliation{National Radio Astronomy Observatory, 520 Edgemont Road, Charlottesville, VA 22903, USA}
\email{}

\author[0000-0003-0489-3750]{Eilat Glikman}
\affiliation{Department of Physics, Middlebury College, Middlebury, VT 05753, USA}
\email{}

\author[0000-0001-7294-106X]{Dong-Chan Kim}
\affiliation{National Radio Astronomy Observatory, 520 Edgemont Road, Charlottesville, VA 22903, USA}
\email{}

\author[0000-0002-1418-3309]{Ji Hoon Kim}
\affiliation{SNU Astronomy Research Center, Department of Physics and Astronomy, Seoul National Univeristy, 1 Gwanak-Ro, Gwanak-Gu, Seoul, 08826, Republic of Korea}
\email{}

\author[0000-0002-3560-0781]{Minjin Kim}
\affiliation{Department of Astronomy, Yonsei University, 50 Yonsei-ro, Seodaemun-gu, Seoul 03722, Republic of Korea}
\email{}

\author[0000-0002-3032-1783]{Mark Lacy}
\affiliation{National Radio Astronomy Observatory, 520 Edgemont Road, Charlottesville, VA 22903, USA}
\email{}

\author[0000-0003-3474-1125]{George C. Privon}
\affiliation{National Radio Astronomy Observatory, 520 Edgemont Road, Charlottesville, VA 22903, USA}
\affiliation{Department of Astronomy, University of Virginia, 530 McCormick Road, Charlottesville, VA 22903, USA}
\affiliation{Department of Astronomy, University of Florida, P.O. Box 112055, Gainesville, FL 32611, USA}
\email{gprivon@nrao.edu}

\author[0000-0002-1568-579X]{Devaky Kunneriath}
\affiliation{National Radio Astronomy Observatory, 520 Edgemont Road, Charlottesville, VA 22903, USA}
\email{}

\author[0009-0002-6248-3688]{Jaya Nagarajan-Swenson}
\affiliation{Department of Astronomy, University of Virginia, 530 McCormick Road, Charlottesville, VA 22903, USA}
\email{}

\author[0000-0003-3638-8943]{N\'uria Torres-Alb\'a}
\affiliation{Department of Astronomy, University of Virginia, 530 McCormick Road, Charlottesville, VA 22903, USA}
\email{}

\author[0009-0002-2049-9470]{Kara N. Green}
\affiliation{Department of Astronomy, University of Virginia, 530 McCormick Road, Charlottesville, VA 22903, USA}
\email{}

\begin{abstract}

We report the results of a pilot study that searched for dual active galactic nuclei (AGN) in local ($z<$\,0.25) galaxies hosting double-peaked narrow emission lines in their optical spectra. We present high-resolution $L-$band (1.5\,GHz or 18\,cm) continuum images from the Very Long Baseline Array (VLBA) as well as WFC3/IR F160W images from the \textit{Hubble Space Telescope} of two candidate dual AGN systems: J0948+6848 and J1223+5409. In both targets, we detected compact non-thermal radio emission that is approximately co-spatial with the near-infrared AGN. Both systems host two high brightness temperature ($>10^{8}$\,K) radio sources that indicate the presence of either a parsec-scale-separation dual AGN ($d_{\text{sep}} \sim 90$\,pc and $\sim 56$\,pc, respectively) or a radio jet. Matched-resolution multi-band radio observations are necessary to further characterize the AGN activity in these systems. 
\end{abstract}

\section{Introduction}

Nearly all galaxies host a supermassive black hole (SMBH). During galaxy mergers, their central SMBHs lose energy and angular momentum via dynamical friction \citep{Chandrasekhar1943, Just2011}, causing their nuclear separations to gradually decrease. Merger-driven inflows of gas and dust fuel accretion onto these SMBHs, triggering intense active galactic nucleus (AGN) activity (e.g, \citealt{Hopkins2006}). These merger-triggered AGN are responsible for the most extreme SMBH accretion events in the Universe (e.g., \citealt{Treister2012, Glikman2015,Weigel2018, Euclid2025}). 

Among the most significant episodes of AGN activity are those in which two (or more) SMBHs are simultaneously active within an interacting system (e.g., \citealt{VW2012}). At kiloparsec and subkiloparsec scale nuclear separations, these systems are referred to as dual AGNs \citep{DeRosa2019, Pfeifle2025}. Binary AGNs, in contrast, are gravitationally-bound Keplerian binaries that form after the merging galaxies have coalesced into a single common envelope \citep{DeRosa2019, Pfeifle2025}. These SMBH binaries occur when nuclear separations have reached parsec to a few tens of parsec scales (depending on the SMBH mass and spin) and will continue to shrink via stellar hardening (e.g., \citealt{Berczik2022}), gas-driven inspirals (e.g., \citealt{Rafikov2016}), and gravitational wave emission (e.g., \citealt{IV2022}). 

The recently detected stochastic gravitational wave (GW) background \citep[e.g.,][]{Agazie2023_PTA, Reardon2023, Xu2023} suggests that merging SMBHs are ubiquitous in the Universe \citep{Agazie2023}. Empirical evidence of such a population, however, is limited. Despite extensive observational surveys at optical, infrared, and X-ray wavelengths (e.g., \citealt{Koss2018, DeRosa2019}), fewer than ten  dual/binary AGNs with subkiloparsec-separations have been confirmed (e.g., \citealt{Fabbiano2011, Kollatschny2020, Koss2023}). In fact, only one
binary AGNs showing two visible nuclei with $<$100\,pc projected nuclear separation has been detected thus far (0402+379;
\citealt{Rodriguez2006}). 


Insufficient spatial resolution and/or dust obscuration are among the factors that are likely responsible for this low detection rate. Specifically, resolving the distinct cores of close-separation AGN pairs often requires parsec-scale spatial resolutions that are only feasible in the local Universe. Furthermore, late-stage mergers, the most probable hosts of dual/binary AGN \citep{Springel2005, Steinborn2016, Volonteri2016, DeRosa2019}, tend to have heavy nuclear obscuration that can render the AGN non-detectable at most wavelengths \citep{Ricci2017, Ricci2021}. Although radio continuum observations remain optically thin up to $N_{H} \sim 10^{26}$ cm$^{-2}$ \citep{Hildebrand1983}, only $\sim10\%$ of AGN are radio-loud \citep{Urry1995, Burke2011, Kellerman2016}. Hence, even extinction-free radio observations would `miss' a significant fraction of AGN in the Universe if not complemented by observations at another wavelength. 

The possibility of false positives in some dual AGN diagnostics further complicates these observational searches. For example, while double-peaked narrow emission lines can be a signature of dual/binary AGNs (e.g., \citealt{Smith2010}), such spectroscopic features are more likely to be tracing gas outflows or rotational motions in the narrow line region \citep{Fue2012}. Similarly, astrometric variability in known AGN could indicate the presence of a hidden close-separation AGN binary (e.g., \citealt{Hwang2020}), but lensed quasar systems (e.g., \citealt{Mannucci2022, Ciurlo2023}), photometric fluctuations due to changes in obscuration, and astrometric errors due to extended host galaxies \citep{Hwang2020} could create similar effects, thus requiring high resolution follow-up observations to confirm dual AGN presence (e.g., \citealt{Chen2023b}). These limitations of single-band observations have emphasized the need for multi-wavelength studies for dual/binary AGN to both maximize the likelihood of detection and provide robust confirmation of an AGN pair.  Indeed, several multi-wavelength studies over the last decade have generated robust dual/binary AGN detections at both low redshift ($z\lesssim0.1$; e.g., \citealt{MS2015, Koss2023, TF2024}) and high redshift ($z\sim2$; e.g., \citealt{Glikman2023, Gross2025}). 

Among these  multi-wavelength identification techniques are searches for positional offsets between radio AGN and optical AGN  (e.g., \citealt{Orosz2013, Kim2019, Veres2021,Chen2023b, Popkov2025,Schwartzman2025}). Specifically, \cite{Kim2019} observed a dual AGN separated by 1.6\,kpc ($0.26''$ at $z=0.3504$) in the galaxy CXO J101527.2+625911. This AGN pair was identified based on the significant spatial offset between the optical AGN and the radio AGN core, indicating that the system must host two distinct AGN.  As further evidence, the [OIII]$\lambda$5007 line emission of the galaxy was best modeled by two emission line components, a phenomenon that can occur when the system hosts two distinct narrow line regions \citep{Kim2017}. It is therefore believed that the system consists of an optically-detected radio-quiet AGN and a radio-detected optically-obscured AGN (otherwise known as a `radio-optical' AGN pair). \cite{Chen2023b} employed a similar method in a systematic study in which they examined 23 dual/binary AGN candidates and compared the coordinates of their Gaia optical positions and their parsec-scale radio AGN emission. They measured significant positional differences in three systems which they identified as candidate dual/binary AGNs with parsec-scale projected nuclear separations. 

The work of \cite{Kim2019} and \cite{Chen2023b} demonstrated that comparisons of radio and optical/near-infrared images could effectively identify dual AGN, even when obscuration is present. In this work, we apply the technique used in \cite{Kim2019} to two candidate dual AGN systems to search for AGN pairs. In Section \ref{sec:sample}, we describe the sample selection criteria, followed by the data description in Section \ref{sec:data}. We present the radio and near-infrared (NIR) images in Section \ref{sec:results} and report the detected AGN positions at each wavelength. In Section \ref{sec:disc}, we discuss the AGN properties and their implications. We present our final conclusions in Section \ref{sec:conclusion}. We adopt a flat cosmology with $H_0 = 70$ km s$^{-1}$ Mpc$^{-1}$, $\Omega_M = 0.3$, and $\Omega_\Lambda = 0.7$. 

\section{Sample}
\label{sec:sample}

\begin{figure*}
    \begin{subfigure}{0.5\textwidth}
      \includegraphics[width=\textwidth]{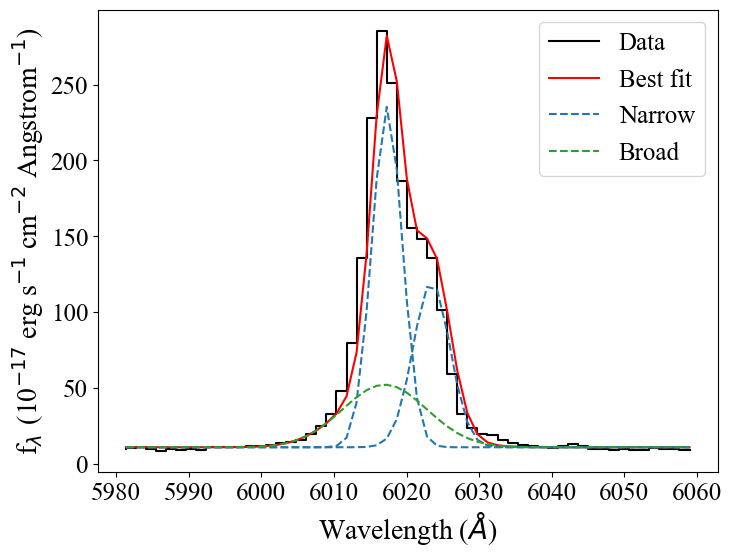}
      \caption{J0948+6848}
    \end{subfigure}
    \hfill
    \begin{subfigure}{0.5\textwidth}
      \includegraphics[width=\textwidth]{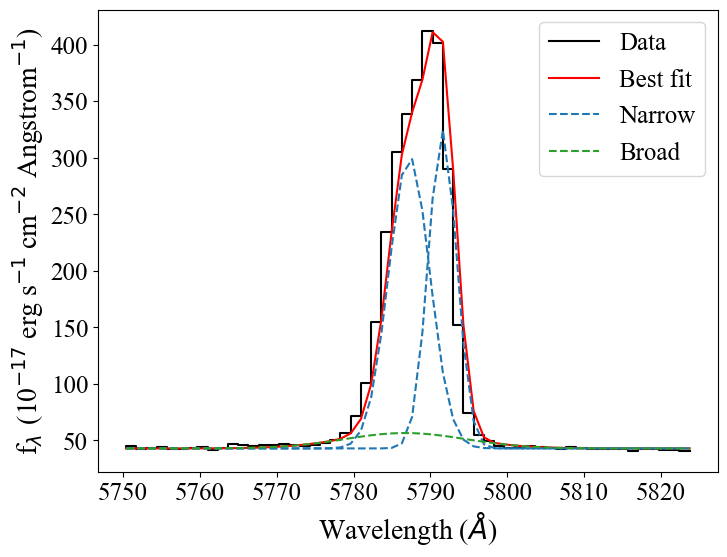}
      \caption{J1223+5409}
    \end{subfigure}
    \vspace{-0.5cm}
    \caption{\textit{SDSS} spectra showing the [OIII]$\lambda$5007 lines of J0948+6848 (left) and J1223+5409 (right). The [OIII] line spectra are fitted best with the double narrow line components (blue) and one broad line component (green). \citet{Kim2020} identified these systems as candidate dual/binary AGN based on these double-peaked [OIII] line profiles.  \label{fig:sdss}}
\end{figure*}

We selected our targets from the complete sample of 1271 local ($z<0.25$) Sloan Digital Sky Survey (SDSS) Data Release 7 (DR7) quasars \citep{Schneider2010}. Spectral decomposition by \cite{Kim2020} found that 77 systems in this parent sample had double-peaked [OIII] emission line profiles. These profiles returned lower reduced $\chi^2$ values when modeled by two Gaussian components rather than a single component, even when a ‘blue-shifted’ broad-line component was included in the model fit (see Figure \ref{fig:sdss} in this work or Table 2 in \citealt{Kim2020}). Such double-peaked line profiles could be tracing a dual/binary AGN system, the rotational kinematics of a single NLR or the interactions of an AGN jet with the interstellar medium \citep{Greene2005, Shen2011, MS2015, Rubinur2019}. The inclusion of the broad-line component in the spectral decomposition, however, helped eliminate double-peaked profiles generated by gas outflows, increasing the likelihood of identifying true dual AGN hosts. 

Of the 77 dual/binary AGN candidates, \cite{Kim2020} found that 27 galaxies in this parent sample were detected at a $>0.72$\,mJy beam$^{-1}$ ($\gtrsim6\sigma$ level, based on the $3\sigma$ contours in their Figure 14) by the first epoch of the Very Large Array Sky Survey (VLASS; \citealt{Gordon2021}; see Figure 14 in \citealt{Kim2020}), indicating the possible presence of at least one radio-bright AGN. We then required that the galaxies have Wide-field Infrared Survey Explorer (WISE) colors W1$-$W2\,$>0.5$ \citep{Assef2013, Satyapal2014}, which reduced the sub-sample to 26 galaxies. We applied this criterion to target galaxy mergers, given that both observations and radiative transfer simulations find that the mid-infrared color rises above this threshold just before the stage of peak black hole growth during galaxy mergers, an epoch during which dual AGN are most likely, making it an effective tool to search for merger-triggered AGN (e.g., \citealt{Lacy2004, Satyapal2014}) and dual AGN hosts \citep{Satyapal2017, Blecha2018}. 

From this carefully selected sample, we target the two best galaxies, SDSS J094822.45+684835.2 ($z$\,=\,0.202) and SDSS J122313.21+540906.5 ($z$\,=\,0.156). These two systems had the brightest VLASS 3\,GHz flux densities presented in \cite{Kim2020} (precise VLASS flux density measurements are presented in Section \ref{sec:disc}) and thus had the highest likelihood of hosting radio AGN.  Hereafter, we refer to the targets as J0948+6848 and J1223+5409, respectively. In this work, we present the first targeted search for dual/binary AGN in these objects. At the redshifts of these galaxies, 1 mas corresponds to $\sim3$\,pc.

\section{Data Description}
\label{sec:data}

\subsection{HST NIR Imaging}

New \textit{Hubble Space Telescope (HST)} images were obtained with the WFC3/IR F160W filter (central wavelength of $\lambda\sim$1.5\,$\mu$m) for both science targets (Program 16485, PI: I. Yoon). J0948+6848 and J1223+5409 were observed on August 27th, 2021 and May 16th, 2022, respectively. The observations were conducted across four identical iterations in a single orbit for a total exposure time of 2623 seconds per object. Data reduction and calibration were performed using the standard \textit{HST} pipeline to produce Hubble Advance Products (HAP) Single Visit Mosaics (SVMs)\footnote[2]{\url{https://archive.stsci.edu/contents/newsletters/december-2020/hap-single-visit-mosaics-now-available}}. These SVM data products were Gaia-aligned \citep{Gaia2023} for improved astrometry and have a pixel-scale of $\sim0.128''/$pixel. The \textit{HST} data presented in this article were obtained from the Mikulski Archive for Space Telescopes (MAST) at the Space Telescope Science Institute. The specific observations analyzed can be accessed via \dataset[doi: 10.17909/y3dt-km68]{https://doi.org/10.17909/y3dt-km68}."

\subsection{VLBA Continuum Imaging}

We obtained high-resolution ($\sim$5 mas) $L$-band (1.5\,GHz or 18\,cm) continuum images using the Very Long Baseline Array (VLBA; Project BY157, PI: I. Yoon). We utilized the polyphase filterbank digital (PFB) system to deliver eight 32 MHz data channels, both with right- and left-hand circular polarizations. This corresponds to a total bandwidth of 256 MHz. The data were recorded at a rate of 2048 Mbit/s with 2-bit sampling. For both targets, we utilized phase-referencing with a three minute cycle time: two minutes on target and one minute on the calibrator. 4C39.25 was used as a fringe finder for both targets and to calibrate the bandpass response. The total on-target observing time was approximately 75 minutes using eight antennas for both science targets. The data were correlated with the DiFX software correlator \citep{Deller2011} in Socorro, New Mexico, with a two-second integration time. The observation details are listed in Table \ref{tab:vlba}.

The data reduction, calibration, and imaging of the data was conducted using the Astronomical Image Processing System (AIPS; \citealt{Greisen2003}), following standard Very Long Baseline Interferometry procedures. Imaging of the phase calibrators confirmed that the emission was compact, allowing for precise phase-referencing calibrations. Self-calibration was performed on both targets to reduce phase and amplitude errors. Since the starting model is the post-phase-referencing model of the target, the additional astrometric uncertainty introduced by the self calibration is estimated to be a fraction of the synthesized beam and does not significantly impact the results presented in this work. The final VLBA continuum images were made with a robust factor of 0 in the AIPS task \texttt{IMAGR}, using the Clark Clean algorithm \citep{Clark1980}. The 1$\sigma$ root-mean-square noises of the continuum images are greater than the theoretical value of ($\sim40\mu$m beam$^{-1}$) estimated by the European VLBI Network observation planner\footnote[3]{\url{https://planobs.jive.eu}} by factors of 1.1 and 1.4 for J0948+6848 and J1223+5409, respectively. We attribute this to significant data loss due to flagging of radio frequency interference ($\sim$40\% and $\sim$38\%, respectively). The synthesized beam sizes and RMS noises of the continuum images are also provided in Table \ref{tab:vlba}. 

\begin{deluxetable}{cccccccc}
\tablecaption{Summary of VLBA L-band (1.5\,GHz) observations\label{tab:vlba}}
\tablewidth{0pt}
\tablehead{\colhead{(1)} &  \colhead{(2)} & \colhead{(3)} & \colhead{(4)} & \colhead{(5)} &  \colhead{(6)} & \colhead{(7)} & \colhead{(8)} \\[-0.2cm]
\colhead{Source} &  \colhead{Date} & \colhead{Phase} & \colhead{Positional Accuracy} & \colhead{Sep} &  \colhead{Beam} & \colhead{P.A.} & \colhead{1$\sigma$ Noise} \\[-0.2cm]
& & \colhead{Calibrator} &  \colhead{(mas)} & \colhead{(deg)} & \colhead{(mas $\times$ mas)} & \colhead{(deg)} & \colhead{($\mu$Jy beam$^{-1}$)}}
\startdata
J0948+6848 & Feb. 01, 2021 & J0958+6533 &  0.03 & 3.40 & 11.5 $\times$ 4.3 & 21.2 & 54 \\
J1223+5409 & Feb. 11, 2021 & J1208+5441 &  0.18 & 2.15 & 11.6 $\times$ 4.1 & 36.6 & 45
\enddata
\tablecomments{Column (1): Source name. Column (2): Date of observations. Column (3): Phase calibrator. Column (4): Positional uncertainty of phase calibrator reported by the VLBA's calibrator search tool\footnote{\url{https://obs.vlba.nrao.edu/cst/}}. Column (5): The angular separation between the phase calibrator and the target in degrees. Column (6): Synthesized beam size of the continuum image. Column (7): Position angle of synthesized beam. Column (8): 1$\sigma$ RMS noise. }
\end{deluxetable}

\section{Results}
\label{sec:results}

\subsection{Modeling the NIR AGN position}

Following the methodology of \citet{Kim2019}, we modeled the host galaxy morphology and NIR AGN position of the \textit{HST} F160W images using the two-dimensional fitting algorithm GALFIT \citep{Peng2010}. In our modeling procedure, we ensured that the fitting region was large enough to include the entire system and sufficient background sky ($0.77'\times0.54'$ and $0.93'\times0.68'$ regions for J0948+6848 and J1223+5409, respectively). We obtained an empirical point spread function (PSF) by extracting images of all stars within the F160W image field of view (2 stars for J0948+6848, 3 stars for J1223+5409) and taking their median. We then modeled the host galaxy with a single PSF, a background sky component, and S\'ersic component(s). Additional S\'ersic components were added to account for neighboring galaxies that overlapped with the host galaxy. All other objects in the field of view were masked in order to reduce the $\chi^2_\nu$ value of the model fit. In the final model, all variables were left as free parameters to ensure unbiased results.

\begin{figure*}
    \centering
    \includegraphics[width=\textwidth]{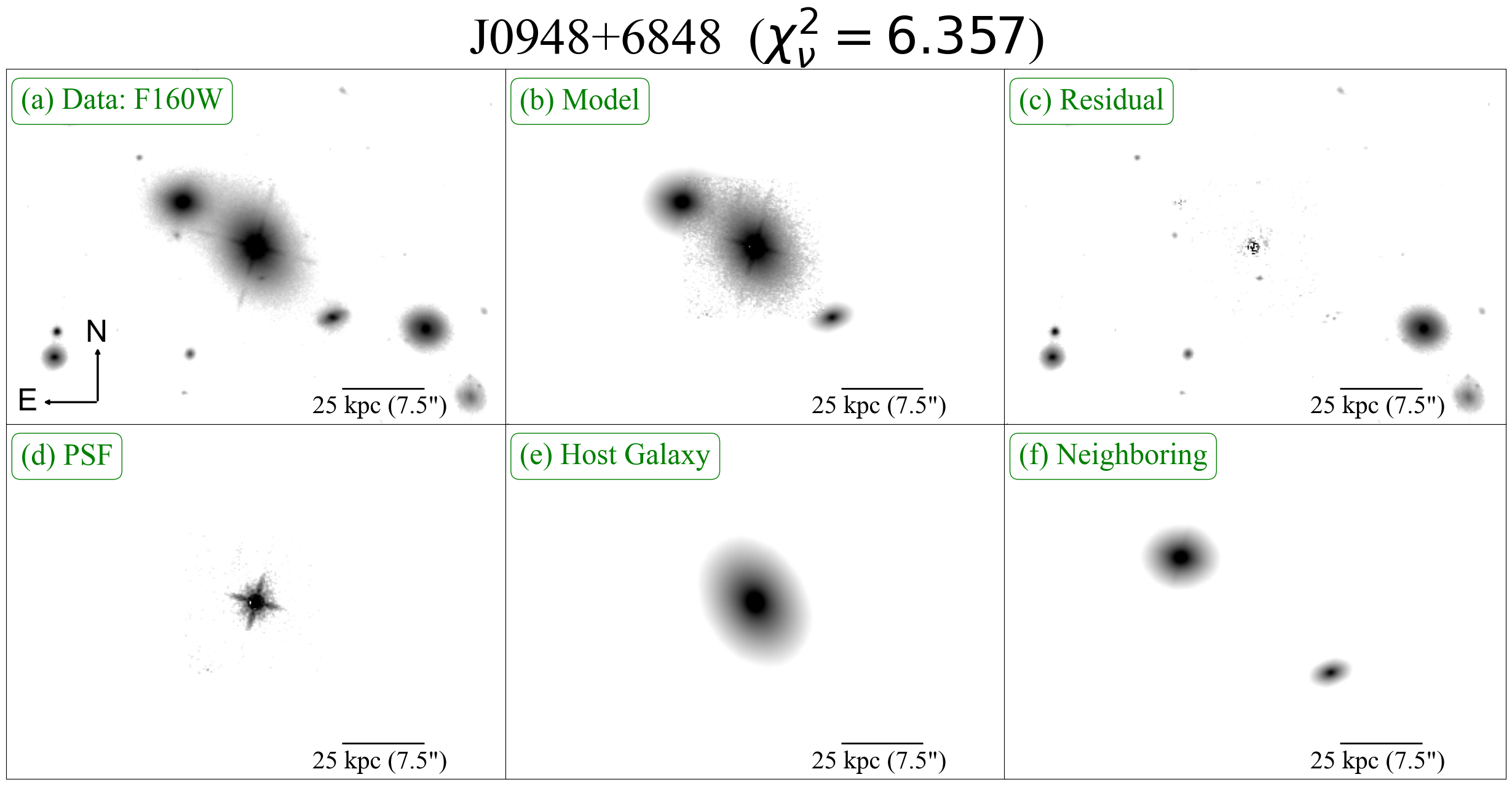}
    \caption{Best-fit GALFIT model of J0948+6848. Top row shows (a) the original image, (b) the model, and (c) the residual image. The bottom row shows the model components: (d) the PSF, (e) the host galaxy (modeled by a single S\'ersic component with index $n\approx1.66$), and (f) the neighboring galaxies. Images are on a logarithmic scale and are $0.77'\times0.54'$ in angular size. The $\chi^{2}_{\nu}$ value of the model fit is reported above the images. We observe a slight artifact of an imperfect PSF which resulted in a higher $\chi^{2}_{\nu}$, but the majority of the flux is well accounted for by the model.   \label{fig:J0948_GALFIT}}
\end{figure*}

\begin{figure*}
    \centering
    \includegraphics[width=\textwidth]{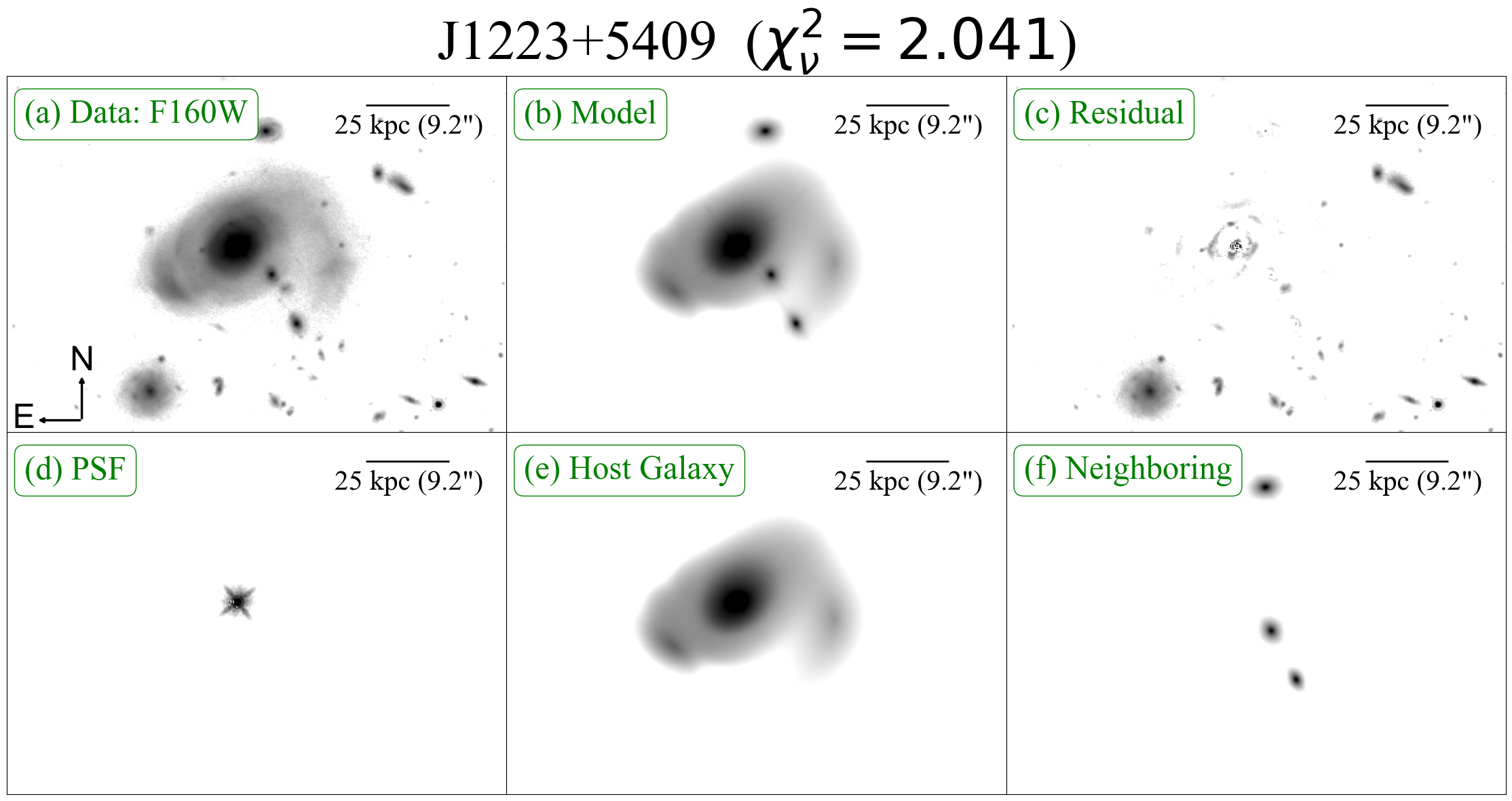}
    \caption{Best-fit GALFIT model of J1223+5409. Top row shows (a) the original image, (b) the model, and (c) the residual image. The bottom row shows the model components: (d) the PSF, (e) the host galaxy (modeled by four S\'ersic components), and (f) the companion galaxies. Images are on a logarithmic scale and are $0.93'\times0.67'$ in angular size. The $\chi^{2}_{\nu}$ value of the model fit is reported above the images.  \label{fig:J1223_GALFIT}}
\end{figure*}

Since J0948+6848 has a relatively symmetric structure, a single S\'ersic component was sufficient to model the central host galaxy (S\'ersic index $n\approx1.66$; Figure \ref{fig:J0948_GALFIT}). Since an exponential disk is defined as having a radial profile with a S\'ersic index of $n=1$, we interpret this as a disk-like galaxy with a slightly perturbed morphology. The galaxy morphology of J1223+5409, in contrast, is highly disturbed with several shell-like features and a tidal tail (Figure \ref{fig:J1223_GALFIT}). We used four S\'ersic components and employed the GALFIT Fourier and bending modes \citep{Peng2010} in addition to the standard S\'ersic parameters. Briefly, Fourier modes add perturbations onto a perfect ellipsoidal shape to create asymmetrical structures. The Fourier perturbations are defined as,
\begin{equation}
    r(x,y) = r_0(x,y) \left( 1 + \sum_{m=1}^{N} a_m \cos(m(\theta+\phi_m))  \right)
\end{equation}
where $r_0(x,y)$ is the radial coordinate of a traditional ellipse, $\theta$ is the position angle, and $a_m$ and $\phi_m$ are the Fourier amplitude and phase angle for mode $m$. Bending modes, in contrast, allow for curvature in the S\'ersic components by perturbing a single axis, following the relation,
\begin{equation}
    y' = y + \sum_{m=1}^{N} a_m \left( \frac{x}{R_{\text{eff}}}\right)^m
\end{equation}
where ($x,y$) are the positions prior to the coordinate transformation, $R_{\text{eff}}$ is the effective radius, and $a_m$ is the amplitude for bending mode $m$.
The addition of these features disentangled the asymmetric and/or distorted structures in J1223+5409 and significantly improved the model fit. We report the `best-fit' GALFIT model parameters in Table \ref{tab:PSF} and present the model images in Figures \ref{fig:J0948_GALFIT} and \ref{fig:J1223_GALFIT}. 

We note that the original \textit{HST} images had brightness units of electrons per second, but we modified the image to be in units of electrons for the GALFIT modeling. This allowed the program to generate correct sigma images to more estimate the $\chi^2_\nu$ value, as explained in the GALFIT advisory document\footnote[4]{https://users.obs.carnegiescience.edu/peng/work/galfit/advisory.html}.  This adjustment, however, resulted in an incorrect magnitude parameter with an offset (the rest of the parameters remain consistent). The magnitudes reported in Table \ref{tab:PSF} have been corrected for this offset and are quoted accurately.

We determined the NIR AGN position to be the PSF position in the `best-fit' GALFIT model (labeled as PSF/AGN in Table \ref{tab:PSF}). We calculated the positional uncertainty of the NIR AGN by summing in quadrature the 1$\sigma$ uncertainty of the GALFIT model, 1$\sigma$ RMS of the right ascension, and  1$\sigma$ RMS of the declination of the Gaia-aligned F160W image. The uncertainties in the Gaia-alignment were taken from FITS file headers of the F160W images (FITS header keywords \texttt{RMS\_RA} and \texttt{RMS\_DEC}) and are automatically generated by the HST pipeline for Gaia-aligned SVM products. For J0948+6848 and J1223+5409, these values were $\sim$8 mas and  $\sim$14 mas for right ascension and $\sim$16 mas and $\sim$8 mas for declination, respectively. The GALFIT contribution to the astrometric uncertainty was minimal, with the outputted PSF positional uncertainty being $<$0.01 pixels ($<$1.28 mas). Combining these contributions, the positional uncertainties of the NIR AGNs are $\sim18$\,mas (J0948+6848) and $\sim16$\,mas (J1223+5409). 


\begin{deluxetable*}{lccccccccccc}
\tablecaption{GALFIT `best fit' model parameters of host galaxies\label{tab:PSF}}
\tabletypesize{\small}
\tablewidth{0pt}
\tablehead{\colhead{(1)} &  \colhead{(2)} & \colhead{(3)} & \colhead{(4)} & \colhead{(5)} & \colhead{(6)} & \colhead{(7)} & \colhead{(8)} & \colhead{(9)} & \colhead{(10)} & \colhead{(11)} \\[-0.2cm]
\colhead{Source}  &  \colhead{Comp.}  &  \colhead{R.A.} & \colhead{Dec.} &  \colhead{Mag} &  \colhead{$R_e$} &  \colhead{$n$} & \colhead{Axis} & \colhead{$\theta$} & \colhead{Fourier} & \colhead{Bending} \\[-0.2cm]
& & \colhead{(J2000)} & \colhead{(J2000)} & \colhead{} & \colhead{($''$)} & & \colhead{Ratio} & \colhead{(deg)} & \colhead{($m$: ($a_m$, $\phi_m$))} & \colhead{($m$: $a_m$)}}
\startdata
J0948+6848 & PSF/AGN & 09:48:$22.463$& +68:48:$35.235$ & $16.45$ &  \\
& S\'ersic & 09:48:$22.462$ & +68:48:$35.228$ & $16.54$ & $2.31$ & $1.66$ & $0.73$ & $29.89$ & \\
\hline
J1223+5409 & PSF/AGN & 12:23:$13.217$ & +54:09:$06.418$ & 17.34 \\
& S\'ersic  & 12:23:$13.215$ & +54:09:$06.445$ & 16.75 & $0.63$ & $1.86$ & $0.80$ & $-66.49$ &  \\
& S\'ersic & 12:23:$13.316$ & +54:09:$05.307$ & $16.78$ & $2.03$ & $0.36$ & $0.82$ & $-54.68$ & 1: (0.79, 30.72) &  \\ 
& S\'ersic & 12:23:$11.952$ &  +54:09:$04.175$ & $17.98$ & $9.89$ & $1.38$ & $0.46$ & $0.81$ & 1: ($-$0.24, $-$100.40) & 2: 35.4 \\
& S\'ersic & 12:23:$14.098$ & +54:09:$01.733$ & $17.00$ & $2.53$ & $0.40$ & $0.94$ & $47.51$ & 1: (0.79, $-$103.26) & 1: $-$4.96; 2: 0.85 \\
\enddata
\tablecomments{Column (1): Source name. Column (2): GALFIT model component. Column (3): Right ascension. The positional uncertainties in right ascension are $\sim$8 mas and $\sim$14 mas for J0948+6858 and J1223+5409, respectively. Column (4): Declination. The positional uncertainty in declination are $\sim$16 mas and $\sim$8 mas for J0948+6858 and J1223+5409, respectively. The positional uncertainties for Columns (3) and (4) were calculated via a quadrature sum of the 1$\sigma$ error outputted by the GALFIT model and 1$\sigma$ error of the Gaia-alignment performed on the HST image. Column (5): Integrated magnitude in the AB system, following the standard definition $m_{tot} = -2.5 \log_{10}(F_{\nu})-48.60$ for a flux density $F_{\nu}$. (6): Effective radius in arcseconds. Column (7): S\'ersic index. Column (8): Ratio of major axis and minor axis. Column (9): Position angle. Column (10): Azimuthal fourier mode, amplitude, and phase angle. Phase angles are reported in degrees. Column (11): Bending mode and amplitude. Note that two different bending modes were used for the one of the S\'ersic components for J1223+5409.}
\end{deluxetable*}


\subsection{Radio properties of the AGN}

\begin{figure*}
    \centering
    \includegraphics[width=0.9\textwidth]{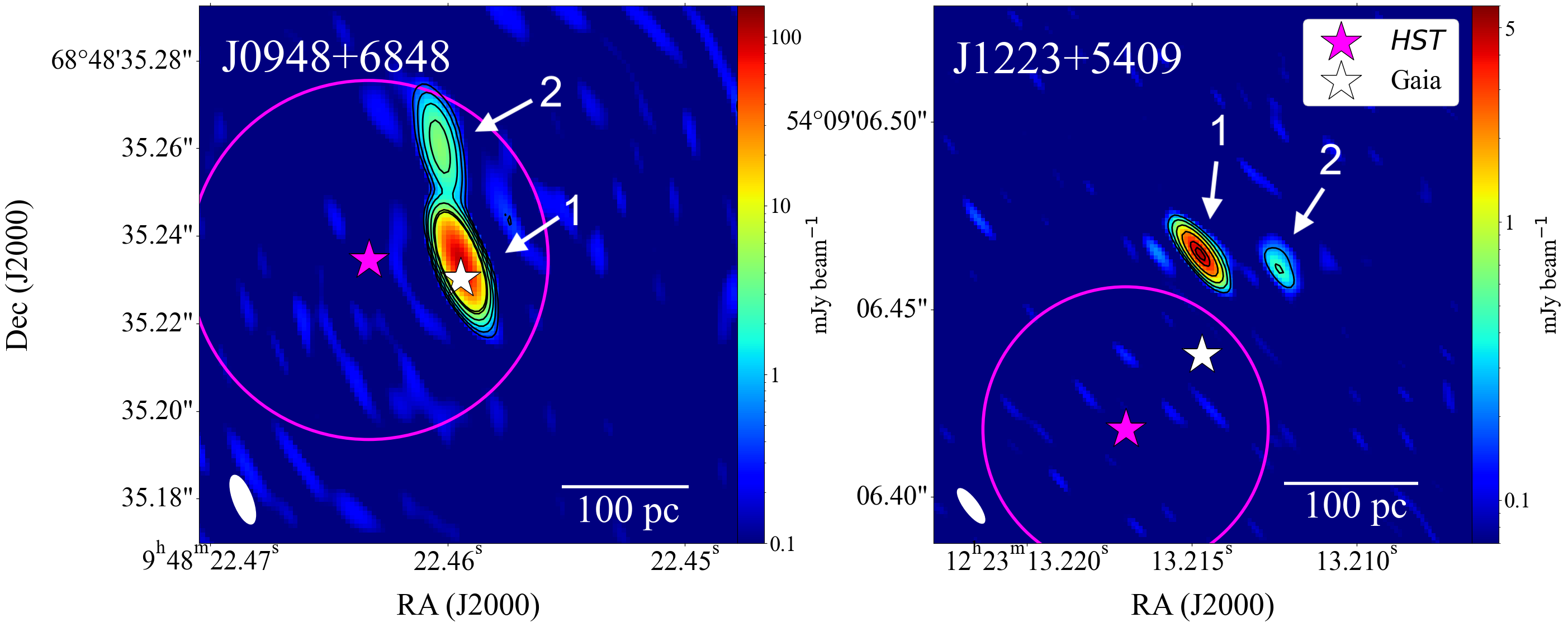}
    \caption{VLBA continuum images at 1.5 GHz for J0948+6848 (left) and J1223+5409 (right) on a logarithmic color scale. Black contours show $5\sigma$, $10\sigma$, $20\sigma$, $40\sigma$, $80\sigma$, $100\sigma$ levels. Detections are labeled as components 1 and 2 in each system. The magenta stars indicate the GALFIT-determined position of the NIR-detected AGN, with the magenta circle representing the 2$\sigma$ relative astrometric uncertainty of the \textit{HST} and VLBA images. The white star is the Gaia-defined position of the host galaxy's nucleus with an uncertainty approximately consistent with the magenta circle. The synthesized beam sizes are shown in the bottom left as white filled ellipses. We do not observe a significant ($>3\sigma$) spatial offset between the NIR-detected AGN and the radio-detected AGN. \label{fig:vlba}}
\end{figure*}

Figure \ref{fig:vlba} shows the 1.5\,GHz continuum images of J0948+6848 and J1223+5409. Both systems have two compact radio continuum sources. We extracted the coordinate positions and radio properties of these detections using the Common Astronomy Software Application's (CASA's) two-dimensional Gaussian fitting tool \texttt{IMFIT} \citep{CASA}. From the derived source sizes and integrated flux densities, we also measured the brightness temperatures of each compact source using the prescription \citep{Condon1992, PT2021},

\begin{equation}
\begin{split}
    T_b & = (1+z)\left(\frac{S_\nu}{\Omega}\right)\frac{c^2}{2kv^2}\\
    & \simeq 1.6\times10^3 \,(1+z)\left(\frac{S_\nu}{\text{mJy}}\right)\left(\frac{\nu}{\text{GHz}}\right)^{-2}\left(\frac{\theta_M\theta_m}{\text{arcsec}^2}\right)^{-1} \text{  K},
\end{split}
\end{equation}
where $c$ is the speed of light, $k$ is the Boltzmann constant, $S_\nu$ is the source flux density measured at frequency $\nu$, and  $\Omega=\pi\theta_M\theta_m$/(4\,ln2) is the solid angle subtended by the source assuming an elliptical Gaussian morphology, with $\theta_M$ and $\theta_m$ corresponding to the FWHM of the deconvolved major and minor axes of the fitted Gaussian, respectively. The factor of ($1+z$) was added to account for the redshift of the source. For unresolved detections, we report brightness temperatures as lower-limits based on the upper-limits on the deconvolved source size. The derived properties are presented in Table \ref{tab:imfit}. The listed coordinates have an astrometric uncertainty of up to $\sim10$ mas \citep{Fomalont1999}, based on the milliarcsecond-scale positional shifts observed by previous VLBA L-band observations using a single phase-referencing calibrator separated by a few degrees from the target (e.g., \citealt{Fomalont1999, Wrobel2004}). The positional uncertainty of the phase calibrator, switching angle errors, and atmospheric variations are likely contributing to these errors.


All four detections have high brightness temperatures ($T_b \sim 10^8-10^{10}$\,K) that are consistent with non-thermal synchrotron-dominated emission driven by AGN activity($>10^5$\,K; e.g., \citealt{Condon1991, Condon1992}). We tentatively interpret the peaks of the radio emission in both target images as the positions of the radio AGN cores. In Section \ref{subsec:disc_J1223}, however, we discuss the possible origins of the secondary detections.

\begin{deluxetable*}{cccccccccc}
\tablecaption{Gaussian-fitted Radio Properties at 1.5\,GHz \label{tab:imfit}}
\tablewidth{0pt}
\tablehead{\colhead{(1)} & \colhead{(2)} & \colhead{(3)} & \colhead{(4)} & \colhead{(5)} & \colhead{(6)} & \colhead{(7)}  & \colhead{(8)} & \colhead{(9)}\\[-0.2cm]
\colhead{Source} & \colhead{Comp.} & \colhead{R.A.} & \colhead{Dec.} & \colhead{$\theta_M$} & \colhead{$\theta_m$} & \colhead{$S_\nu$} & \colhead{Peak Flux}  & \colhead{$T_b$}\\[-0.2cm]
\colhead{} & \colhead{} & \colhead{(J2000)} & \colhead{(J2000)} & \colhead{(mas)} & \colhead{(mas)} &  \colhead{(mJy)} & \colhead{(mJy  beam$^{-1}$)} &  \colhead{(K)} }  
\startdata
J0948+6848 & 1 & 09:48:22.460 & +68:48:35.234  & 3.0$\pm$0.1 & 0.55$\pm$0.06 & 106.7$\pm$11 & 102.3$\pm$10 (1894$\sigma$)  & (5.4$\pm$0.8)$\times$10$^{10}$ \\
& 2 & 09:48:22.460 & +68:48:35.258 & 19.6$\pm$3.3 & 1.6$\pm$1.1 & 9.2$\pm$1.3 & 3.9$\pm$0.6 (72$\sigma$) & (2.5$\pm$1.8)$\times$10$^{8}$ \\
J1223+5409 & 1 & 12:23:13.215 & +54:09:06.465 & $<2.0^*$ & $<0.94^*$ & 5.1$\pm$0.5 & 5.0$\pm$0.5 (111$\sigma$) & $\geq$\,2.2$\times$10$^{9}$ \\
& 2 & 12:23:13.212 & +54:09:06.461 & $<2.0^*$ & $<0.94^*$ & 0.7$\pm$0.1 & 0.45$\pm$0.06 (10$\sigma$) & $\geq$\,3.1$\times$10$^{8}$
\enddata
\tablenotetext{*}{Upper-limits on the deconvolved source size due to unresolved point source. }
\tablecomments{Column (1): Source name. Column (2): Component number as labeled in Figure \ref{fig:vlba}. Column (3): Right ascension of the detection. Column (4): Declination of the detection. For the coordinate positions in columns (3) and (4), the astrometric uncertainty is $\sim$10 mas. Columns (5) and (6): FWHM of the major and minor axes of the radio detection, deconvolved from the synthesized beam.  Column (7): Integrated flux density at 1.5\,GHz (18\,cm). Column (8): Peak flux density at 1.5\,GHz (18\,cm).  The sigma levels in parentheses refer to the significance of the measured flux density with respect to the 1$\sigma$ RMS noise of the image. Column (9): Redshift-corrected brightness temperature. The 1$\sigma$ uncertainties listed for Columns (5) through Column (9) are based on the estimated flux calibration uncertainty ($\sim$10\%) and/or the \texttt{IMFIT} parameter uncertainties. }
\end{deluxetable*}

\subsection{Radio-Near Infrared AGN offsets}

\begin{figure*}
    \centering
    \includegraphics[width=0.9\textwidth]{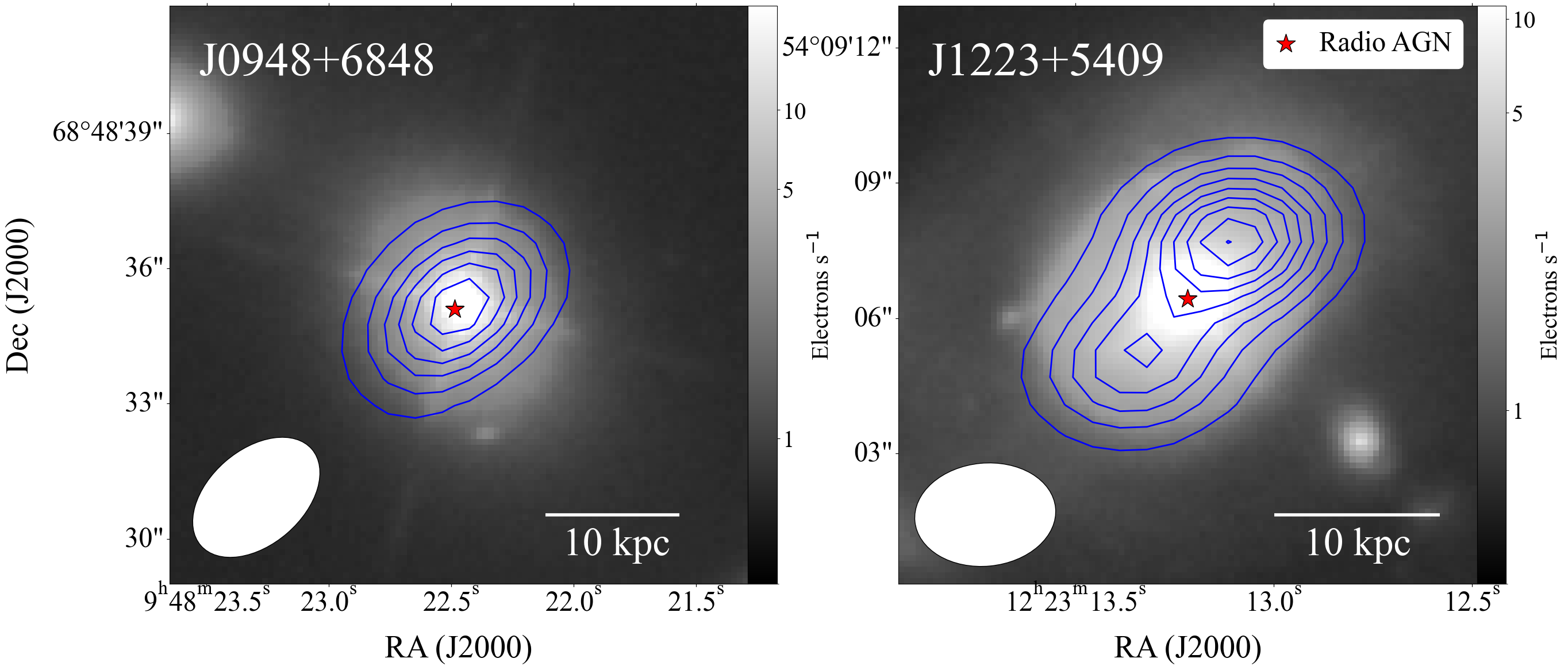}
    \caption{Greyscale WFC3/IR F160W images of J0948+6848 (left) and J1223+5409  (right) on a logarithmic scale. Images are oriented such that north is up and east is to the left and are $\sim$0.21$'\times0.21'$ in angular size.  The red star is the position of the brightest VLBA-detected radio source, though the position of the secondary source also falls within the area of the star. VLASS 3\,GHz radio contours are overlaid in blue, showing $100\sigma$ levels (100$\sigma$, 200$\sigma$, 300$\sigma$....) where the 1$\sigma$ RMS is $\sim120\mu$Jy beam$^{-1}$. The synthesized radio beam is shown as a white ellipse in the bottom left. For both systems, the VLBA-detected AGN cores and NIR center are relatively spatially coincident. The VLASS radio peak of J1223+5409, however, is offset by kiloparsec-scales. \label{fig:hst_VLASS}}
\end{figure*}

Here, we compare the NIR AGN position (Table \ref{tab:PSF}) and the radio AGN position (Table \ref{tab:imfit}). We sum in quadrature the positional uncertainties of the NIR AGN positions ($\sim18$\,mas and $\sim16$\,mas for J0948+6848 and J1223+5409, respectively) and the VLBA positional uncertainty ($\sim10$\,mas) to calculate $1\sigma$ relative astrometric uncertainties of $\sim$20\,mas and $\sim$19\,mas, respectively. We find that the projected spatial offsets are $<2\sigma$ for J0948+6848 and $<3\sigma$ for J1223+5409, and are thus insignificant. This places a 3$\sigma$ upper-limit of $<$268\,pc ($<$0.08$''$)and $<$297\,pc ($<$0.11$''$) (calculated by summing the measured projected offset and the $3\sigma$ relative astrometric uncertainties), respectively, for the projected spatial offset between the radio AGN and NIR AGN \textit{if} the systems host `radio-NIR' dual AGN similar to CXO J101527.2+625911 \citep{Kim2019}. We tentatively conclude that the radio-detected and NIR-detected AGN are spatially coincident.

We also compare the AGN positions to the optical nucleus of each system (marked as a white star in Figure \ref{fig:vlba}), as reported in the Gaia Data Release 3 catalog \citep{Gaia2023}. We find that the Gaia positions are within the $3\sigma$ astrometric uncertainty of both the NIR and radio AGN positions. We therefore deduce a tentative positional agreement between the optical, NIR, and radio detections, despite the higher rates of error in Gaia's astrometric solutions for extended host galaxies, particularly at lower red-shift \citep{Hwang2020}.

\section{Discussion}
\label{sec:disc}

While we cannot confirm the presence of a `radio-NIR' dual AGN in our pilot sample, the detection of two high brightness temperature ($>10^8$\,K; Table \ref{tab:imfit}) radio continuum detections in both systems suggest two possible scenarios: a dual radio AGN scenario and a radio core-jet scenario. Since our single radio band observations are insufficient to derive a precise spectral index to distinguish between flat-spectrum AGN cores (e.g., \citealt{Nagar2000}) and steep-spectrum radio jets (e.g., \citealt{Hovatta2014}), we instead utilize archival data sets to consider each case.

\subsection{Dual/Binary AGN scenario}

In a dual/binary AGN scenario, the pairs of radio detections would represent two distinct AGN cores with projected nuclear separations of $\sim$\,90\,pc and $\sim$\,56\,pc for J0948+6848 and J1223+5409, respectively. These projected nuclear separations are comparable to those estimated for the candidate dual/binary AGN identified based on Gaia-VLBA offsets by \cite{Chen2023b}. At these projected separations, the SMBH pair is likely gravitationally bound and is expected to be in the stellar hardening phase of the SMBH merger in which the SMBH pair sheds energy primarily via stellar three-body interactions \citep{Koss2023}.

For each system, we estimate the projected velocity difference of the co-orbiting SMBHs from the spectral decomposition of the double-peaked [OIII] lines reported by \citet{Kim2020} (278\,km s$^{-1}$ and 232\,km s$^{-1}$ for J0948+6848 and J1223+5409, respectively). Assuming a Keplerian orbit, the sum of the two SMBH masses would be log($M_{\text{BH, total}}/M_\odot$) $\sim$ 9.2 and 8.8, respectively. Dual AGN with SMBH masses of similar orders of magnitude have been detected (e.g., log($M_{\text{BH, total}}/M_\odot$) $\sim$ 8.6 for UGC 4211; \citealt{Koss2023}), making this a plausible scenario. 

We note, however, that this calculation assumes two distinct NLRs, each associated with a SMBH. Indeed, in a previous study, only $\sim$2\% of double-peaked [OIII] lines are attributed to dual/binary AGN with separate NLRs \citep{Fue2012}. Since the [OIII] emission discussed here is not spatially resolved to the parsec-scales probed by the VLBA, we cannot prove nor disprove this scenario.

\subsection{Core-Jet scenario}
\label{subsec:disc_J1223}



Alternatively, it is also possible that our VLBA detections are tracing clumpy emission from a radio jet. As determined from the relative positions of the two compact VLBA detections, such a core-jet structure would have a position angle of $\sim$10 degrees and $\sim$91 degrees for J0948+6848 and J1223+5409, respectively (rotating east where north is 0 degrees).

For J1223+5409, this appears to conflict with the position angle of the potential radio jet detected at lower resolutions. For example, \cite{Jarvis2021} conducted multi-frequency  (1.4\,GHz and 5\,GHz) radio observations of J1223+5409 at $\sim0.3''$ resolution ($\sim$800 parsecs) using the Very Large Array (VLA). They constrained a spectral index of approximately $\alpha_{1.4}^{5.0} = -0.5 \pm 0.3$ for the core (assuming $S_\nu \propto \nu^{\alpha}$ where $S_\nu$ is the flux density at frequency $\nu$), which they tentatively report as evidence of an AGN with an active radio jet. Both these VLA observations by \citet{Jarvis2021} (see their Figure 4) and the lower resolution 3\,GHz VLASS images of J1223+5409 (see blue contours on the right panel of Figure \ref{fig:hst_VLASS}) reveal a bipolar emission structure extending from the central core, possibly tracing a large-scale radio jet. From the VLASS image, we estimate that the structure is elongated along an axis with a 21 degree position angle. Given the difference of $\sim70$ degrees compared to the VLBA-determined position angle, this suggests that, in a core-jet scenario, the kiloparsec-scale jet structure has a drastically different orientation than the parsec-scale jet emission.

\begin{deluxetable}{ccc}[h]
\tablecaption{VLASS Flux Densities at 3\,GHz \label{tab:vlass}}
\tablewidth{0pt}
\tablehead{\colhead{(1)} & \colhead{(2)} & \colhead{(3)} \\[-0.2cm]
\colhead{Source} & \colhead{Date} & \colhead{$S_{3\text{GHz}}$} \\[-0.2cm]
\colhead{} & \colhead{} & \colhead{(Jy)} }  
\startdata
J0948+6848 & Oct. 21, 2017 & 0.090 $\pm$ 0.009 \\
&  Sep. 01, 2020 & 0.106 $\pm$ 0.003 \\ 
& Feb. 15, 2023 & 0.105 $\pm$ 0.003 \\
\hline
J1223+5409 & Nov. 17, 2017 & 0.181 $\pm$ 0.019 \\
& Aug. 20, 2020 & 0.204 $\pm$ 0.006\\
& Feb. 7, 2023 & 0.200 $\pm$ 0.006 \\
\enddata
\tablecomments{Column (1): Source name. Column (2): Date of VLASS observations. Column (3): Total flux density at 3\,GHz, measured over a circular aperture with a 20$''$ radius. The reported uncertainty considers 1$\sigma$ RMS of the image and total flux density uncertainty of VLASS ($\sim10\%$ for epoch 1.1, and $\sim3\%$ for the remaining epochs).}
\end{deluxetable}

Similar systems have been observed with different radio jet trajectories at parsec-scales compared to kiloparsec-scales (e.g., \citealt{Beswick2004}), pointing to the possibility of extreme jet axis precession and/or the formation of a new radio jet in these systems. Newborn radio jets, however, are expected to generate high flux variability over year to decade timescales (100$-$2500\%, \citealt{Nyland2020, Wolowska2021, Zhang2022}). We searched for evidence of this flux variability by comparing VLASS 3\,GHz continuum images across the three observing epochs. To maximize image quality, we utilized the publicly-available calibrated measurement sets and performed self-calibration on both the phase calibrator and the final target using the customized VLASS Single Epoch Continuum Pipeline \citep{VLASS_pip}. Extracting flux densities using a circular aperture with a 20$''$ radius, we measured only a $\sim$10$-$15\% flux density increase over the three observing epochs in both systems (Table \ref{tab:vlass}). Although this is tentatively greater than the estimated flux density uncertainty of the calibrated images ($\sim$3\%\footnote[5]{\url{https://library.nrao.edu/public/memos/vla/vlass/VLASS_013.pdf}}), the measured flux density change could be a product of the antenna position error during the first epoch of VLASS that led to systematically low total flux density measurements (differences as high as 10\%; \citealt{Gordon2021}). It is therefore unclear whether this flux density difference is physically meaningful.

Future matched-resolution multi-band radio observations will allow us to constrain spectral indices for each detection and differentiate between these two scenarios. Follow-up study, however, is warranted in either scenario given that these systems either (1) host some of the closest separation dual AGN detected thus far or (2) have experienced a sudden reorientation of the AGN jet-axis due to merger-driven dynamics (e.g., \citealt{Roos1988, Abraham2018}) or accretion instabilities (e.g., \citealt{Liska2018, Lalakos2022}).


\section{Conclusion}
\label{sec:conclusion}

Using high-resolution NIR and radio images from the \textit{HST} and the VLBA, we searched for AGN pairs in two local ($z<0.25$) candidate dual AGN \citep{Kim2020}. From the NIR images, we modeled the galaxies' structure to constrain the NIR AGN position. From the radio continuum images, we performed two-dimensional Gaussian fitting to extract the source position and radio properties. Our results are as follows:

\begin{itemize}
    \item In both targets, we do not find significant spatial offsets between the NIR-detected AGN and the radio-detected AGN. If the AGN observed at each wavelength are distinct, then their projected nuclear separation must be $<$268\,pc ($<$0.08$''$) and $<$297\,pc ($<$0.11$''$) for J0948+6848 and J1223+5409, respectively.


    \item We detected two compact radio sources in both J0948+6848 and J1223+5409 that have brightness temperatures consistent with AGN activity ($>10^7$\,K). The galaxies host either a pair of radio AGN cores or a core-jet system with clumpy emission from a radio jet. Future matched-resolution multi-band radio observations are necessary to distinguish between these scenarios. 
\end{itemize}

\begin{acknowledgments}
The National Radio Astronomy Observatory is a facility of the National Science Foundation operated under cooperative agreement by Associated Universities, Inc. This research is based on observations made with the NASA/ESA Hubble Space Telescope obtained from the Space Telescope Science Institute, which is operated by the Association of Universities for Research in Astronomy, Inc., under NASA contract NAS 5–26555. These observations are associated with program HST-GO-16485. MAJ acknowledges support from grant HST-GO-16485.001-A. This work made use of the Swinburne University of Technology software correlator, developed as part of the Australian Major National Research Facilities Programme and operated under license. MK was supported by the National Research Foundation of Korea (NRF) grant funded by the Korean government (MSIT) (Nos. RS-2024-00347548 and RS-2025-16066624). JHK acknowledges the support from the National Research Foundation of Korea (NRF) grants, No. 2021M3F7A1084525 and No. 2020R1A2C3011091, and the Institute of Information \& Communications Technology Planning \& Evaluation (IITP) grant, No. RS-2021-II212068 funded by the Korean government (MSIT). EG acknowledges the generous support of the Cottrell Scholar Award through the Research Corporation for Science Advancement. EG is grateful to the Mittelman Family Foundation for their generous support.

\end{acknowledgments}

\bibliography{bib}{}
\bibliographystyle{aasjournal}

\end{document}